\documentclass[prl,superscriptaddress,twocolumn]{revtex4}

\usepackage{amsfonts,amssymb,amsmath}
\usepackage[]{graphics,graphicx,epsfig}
\usepackage{amsthm}

\def\identity{\leavevmode\hbox{\small1\kern-3.8pt\normalsize1}}

\newcommand{\ket}[1]{\left | #1 \right\rangle}
\newcommand{\bra}[1]{\left \langle #1 \right |}

\newcommand{\Tr}{\mathrm{Tr}}

\renewcommand{\epsilon}{\varepsilon}

\begin{document}

\title{Correlation complementarity yields Bell monogamy relations}

\author{P. Kurzy\'nski}
\affiliation{Centre for Quantum Technologies, National University of Singapore, 3 Science Drive 2, 117543 Singapore, Singapore}
\affiliation{Faculty of Physics, Adam Mickiewicz University, Umultowska 85, 61-614 Pozna\'{n}, Poland}

\author{T. Paterek}
\affiliation{Centre for Quantum Technologies, National University of Singapore, 3 Science Drive 2, 117543 Singapore, Singapore}

\author{R. Ramanathan}
\affiliation{Centre for Quantum Technologies, National University of Singapore, 3 Science Drive 2, 117543 Singapore, Singapore}

\author{W. Laskowski}
\affiliation{Institute of Theoretical Physics and Astrophysics, University of Gda\'nsk, 80-952 Gda\'nsk, Poland}
\affiliation{Fakult\"at f\"ur Physik, Ludwig-Maximilians Universit\"at M\"unchen, 80755 M\"unchen, Germany}
\affiliation{Max Planck Institut f\"ur Quantenoptik, 85748 Garching, Germany}

\author{D. Kaszlikowski}
\email{phykd@nus.edu.sg}
\affiliation{Centre for Quantum Technologies, National University of Singapore, 3 Science Drive 2, 117543 Singapore, Singapore}
\affiliation{Department of Physics, National University of Singapore, 2 Science Drive 3, 117542 Singapore, Singapore}

\begin{abstract}
We present a method to derive Bell monogamy relations by connecting the complementarity principle with quantum non-locality. The resulting monogamy relations are stronger than those obtained from the no-signaling principle alone. In many cases, they yield tight quantum bounds on violation of single and multiple qubit correlation Bell inequalities. In contrast with the two-qubit case, a rich structure of possible violation patterns is shown to exist in the multipartite scenario. 
\end{abstract}

\maketitle

It is an experimentally confirmed fact, with the exception of certain experimental loopholes, that Bell inequalities are violated \cite{BELL}.
In a typical Bell scenario a composite system is split between many parties and each party independently performs measurements on their corresponding subsystems.
When all measurements are done the parties meet and calculate a function (Bell parameter) of their measurement outcomes in order to check whether they succeeded in violation of local realism.
An interesting phenomenon occurs when a subsystem is involved in more than one Bell experiment, i.e. when measurement outcomes of one party are plugged into more than one Bell parameter involving different parties.
In this case trade-offs exist between strengths of violations of a Bell inequality by different sets of observers, known as monogamy relations \cite{SG2001,TV2006,BLMPSR2005,MAN2006,TONER2009,PB2009}.
One of the origins of this monogamy is the principle of no-signaling,
according to which information cannot be transmitted with innite speed.
If violations are sufficiently strong possibility of superluminal communication between observers arises and consequently the Bell monogamy is present in every no-signaling theory \cite{BLMPSR2005,MAN2006,TONER2009,PB2009}.
However, no-signaling principle alone does not identify the set of violations allowed by quantum theory.
The monogamy relations derived within quantum theory, in the scenario where a Bell inequality
is tested between parties $AB$ and $AC$, show even more stringent constraints on the allowed violations \cite{SG2001,TV2006}.

Here we derive within quantum theory the monogamy relations which involve violation of multi-partite Bell inequalities, and study their properties.
The trade-offs obtained are stronger than those arising from no-signaling alone
and in most cases we show that they fully characterize the quantum set of allowed Bell violations.
Our method uses complementarity of operators defining quantum values of Bell parameters
and shows that Bell monogamy stems from quantum complementarity.

This sheds new light on the relation between complementarity (uncertainty) and quantum non-locality.
Oppenheim and Wehner show that complementarity relations for single-party observables determine the strength of a single Bell inequality violation \cite{OW}.
Here we show for qubit inequalities that the same can be achieved using complementarity between correlation observables
and that this type of complementarity also determines violation strength for several Bell inequalities (monogamy).

We begin with the principle of complementarity, which forbids simultaneous knowledge of certain observables, and show that the only dichotomic complementary observables in quantum formalism are those that anti-commute. 
Conversely, we demonstrate that there exists a bound for the sum of squared expectation values of anti-commuting operators in any physical state \cite{GEZA,WW2008}.
This bound is subsequently used to derive quantum bounds on Bell inequality violations.
For its other applications see for instance Ref. \cite{MACRO}.

Consider a set of dichotomic ($\pm 1$) complementary measurements. The complementarity is manifested in the fact that if the expectation value of one measurement is $\pm1$ then expectation values of all other complementary measurements are zero. We show that the corresponding quantum mechanical operators anti-commute.
Consider a pair of dichotomic operators $A$ and $B$ and put the expectation value $\langle A \rangle = 1$, i.e., the state being measured, say $\ket{a}$, is one of the $+1$ eigenstates.
Complementarity requires $\bra{a} B \ket{a} = 0$, which implies $B \ket{a} = \ket{a_{\perp}}$, where $\perp$ denotes a state orthogonal to $\ket{a}$.
Since $B^2 = \openone$, we also have $B \ket{a_{\perp}} = \ket{a}$ and therefore $\ket{b} = \tfrac{1}{\sqrt{2}}(\ket{a} + \ket{a_\perp})$ is the $+1$ eigenstate of $B$.
For this state complementarity demands, $\bra{b} A \ket{b} = 0$, i.e. $A \ket{b}$ is orthogonal to $\ket{b}$ which is only satisfied if $\ket{a_\perp}$ is the $-1$ eigenstate of $A$.
The same argument applies to all $+1$ eigenstates, therefore the two eigenspaces have equal dimension.
As a consequence, $A = \sum_{a} (\ket{a} \bra{a} - \ket{a_\perp} \bra{a_\perp})$ and $B = \sum_{a} (\ket{a_{\perp}} \bra{a} + \ket{a} \bra{a_{\perp}})$.
It is now easy to verify that $A$ and $B$ anti-commute.

Conversely, consider a set of traceless and trace-orthogonal dichotomic hermitian operators $A_k$.
We denote by $\alpha_k$ the expectation values of measurements  $A_k$ in some state $\rho$, which are real numbers in the range $[-1,1]$. 
Let us group operators $A_k$ into disjoint sets $S_j$ of mutually anti-commuting operators, $S_j=\{A_1^{(j)},A_2^{(j)},\dots\}$.
Next, consider an operator $F_j \equiv \sum_{k=1}^{|S_j|} \alpha_{kj} A_{k}^{(j)} = \vec \alpha_j \cdot \vec A_j$, whose variance in the same state $\rho$ is given by $\langle F_j^2\rangle-\langle F_j\rangle^2 = |\vec \alpha_j|^2 (1- |\vec \alpha_j|^2)$
due to assumed anti-commutativity and because the square of each individual operator is identity.
Positivity of variance, which stems from the positivity of $\rho$, implies that
\begin{equation}
|\vec\alpha_{j}|\leq 1.
\label{INEQ}
\end{equation}
As a result, if an expectation value of one observable is $\pm 1$ then expectation values of all other anti-commuting observables are necessarily zero.
In this way anti-commuting operators are related to complementarity.
In fact, the above inequality is more general as it gives trade-offs between squared expectation values of anti-commuting operators in any physical state.

Here, we derive inequality (\ref{INEQ}) in the spirit of Heisenberg uncertainty relation, see \cite{GEZA,WW2008} for alternative derivations. For dichotomic observables the square of
expectation value is related to the Tsallis entropy as $S_2(A_j)=\frac{1}{2}(1-\langle A_j\rangle^2)$,
therefore the inequality can be converted into entropic uncertainty relation.

Inequality (\ref{INEQ}) provides a powerful tool for the studies of quantum non-locality.
We show that it allows derivation of the Tsirelson bound \cite{TSIRELSON} and monogamy of Bell inequality violations between many qubits.
A general $N$-qubit density matrix can be decomposed into tensor products of Pauli operators
\begin{equation}
\rho = \frac{1}{2^N} \sum_{\mu_1,...,\mu_N =0}^3 T_{\mu_1 \dots \mu_N} \sigma_{\mu_1} \otimes \dots \otimes \sigma_{\mu_N},
\end{equation}
where $\sigma_{\mu_n} \in \{\openone, \sigma_x,\sigma_y,\sigma_z\}$ is the $\mu_n$-th local Pauli operator for the $n$-th party and $T_{\mu_1 \dots \mu_N} = \Tr[\rho (\sigma_{\mu_1} \otimes \dots \otimes \sigma_{\mu_N}) ]$ are the components of the correlation tensor $\hat T$.
The orthogonal basis of tensor products of Pauli operators has the property that its elements either commute or anti-commute.

We study a complete collection of two-setting correlation Bell inequalities for $N$ qubits \cite{WZ2001,WW2001,ZB2002}.
It can be condensed into a single general Bell inequality, whose classical bound is one \cite{ZB2002}.
All correlations which satisfy this general inequality and only such correlations admit a local hidden variable (LHV) description of the Bell experiment.
This is in contrast to single inequality like, e.g., CHSH \cite{CHSH} violation of which is only sufficient to disqualify LHV model.
For two qubits, if the general inequality is satisfied then all CHSH inequalities are satisfied,
and if the general inequality is violated then there exists a CHSH inequality (with minus sign in a suitable place) which is violated.
The quantum value of the general Bell parameter, denoted by $\mathcal{L}$, was shown to have an upper bound of
\begin{equation}
\mathcal{L}^2 \le \sum_{k_1, \dots, k_N=x,y} T_{k_1 \dots k_N}^2,
\label{UP_BOUND}
\end{equation}
where summation is over orthogonal local directions $x$ and $y$ which span the plane of the local settings \cite{ZB2002}. If the upper bound above is smaller than the classical limit of $1$, there exists an LHV model. 
Our method for finding quantum bounds for Bell violations is to use condition (\ref{UP_BOUND})
for combinations of Bell parameters and then identify sets of anti-commuting operators in order to utilize inequality (\ref{INEQ}) and obtain a bound on these combinations.

We begin by showing an application of Inequality (\ref{INEQ}) to a new derivation of the Tsirelson bound.
For two qubits the general Bell parameter is upper bounded by $\mathcal{L}^2 \le T_{xx}^2 + T_{xy}^2 + T_{yx}^2 + T_{yy}^2$.
One can identify here two vectors of averages of anti-commuting observables, e.g., $\vec \alpha_1 = (T_{xx},T_{xy})$ and $\vec \alpha_2 = (T_{yx},T_{yy})$.
Due to (\ref{INEQ}) we obtain $\mathcal{L} \le \sqrt{2}$ which is exactly the Tsirelson bound.
One can apply this method to look for corresponding maximal quantum violations of other correlation inequalities,
e.g. it is easy to verify that the ``Tsirelson bound'' of the multi-setting inequalities \cite{MULTISETTING} is just the same as the one for the two-setting inequalities. Our derivation shows that Tsirelson's bound is due to complementarity of correlations $T_{ix}^2+T_{iy}^2 \leq 1$ with $i=x,y$.
Any theory more non-local than quantum mechanics would have to violate this complementarity relation (compare with Ref. \cite{OW}).

\begin{figure}
\begin{center}
\includegraphics[scale=0.4]{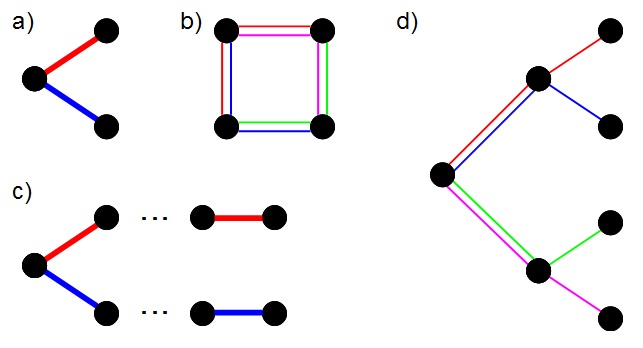}
\end{center}
\caption{The nodes of these graphs represent observers trying to violate Bell inequalities which are denoted by colored edges.
{\bf a)} The simplest case: two subsets of three parties try to violate CHSH inequality.
{\bf b)} Four three-party subsets of four parties try to violate Mermin inequality.
{\bf c)} Two subsets of odd number of parties try to violate multi-partite Bell inequality in a scenario in which only one particle is common to two Bell experiments.
{\bf d)} A binary tree configuration leads to strong monogamy relation.
}
 \label{FIG_GRAPHS}
\end{figure}

To describe how complementarity of correlations can be used to establish Bell monogamy, consider the simplest scenario of three particles, illustrated in Fig. \ref{FIG_GRAPHS}a.
We show that if correlations obtained in two-setting Bell experiment by $AB$ 
cannot be modeled by LHV, then correlations obtained by $AC$ admit LHV model.
We use condition (\ref{UP_BOUND}) which applied to the present
bipartite scenario reads: $\mathcal{L}_{AB}^2 + \mathcal{L}_{AC}^2 \le \sum_{k,l=x,y} T_{kl0}^2 + \sum_{k,m=x,y} T_{k0m}^2$.
It is important to note that the settings of $A$ are the same in both sums
and accordingly orthogonal local directions $x$ and $y$ are the same for $A$ in both sums.
We arrange the Pauli operators corresponding to correlation tensor components entering the sums 
into the following two sets of anti-commuting operators: $\{XX\openone,XY\openone,Y\openone X,Y\openone Y\}$ and $\{YX\openone,YY\openone, X\openone X,X \openone Y\}$, where $X=\sigma_x$ and $Y=\sigma_y$.
Note that the anti-commutation of any pair of operators within a set is solely due to anti-commutativity of local Pauli operators.
We obtain our result $\mathcal{L}_{AB}^2 + \mathcal{L}_{AC}^2 \le 2$. 
Once a CHSH inequality is violated between $AB$, all CHSH inequalities between $AC$ are satisfied, similar results were obtained in \cite{SG2001,TV2006}.

Before we move to a general case of arbitrary number of qubits, we present an explicit example of multipartite monogamy relation.
Consider parties $A$, $B$, $C$, $D$ trying to violate a correlation Bell inequality in a scenario depicted in Fig. \ref{FIG_GRAPHS}b.
We show the new monogamy relation:
$\mathcal{L}_{ABC}^2+\mathcal{L}_{ABD}^2+\mathcal{L}_{ACD}^2+\mathcal{L}_{BCD}^2 \leq 4$.
Condition (\ref{UP_BOUND}) applied to these tripartite Bell parameters implies that
the left-hand side is bounded by the sum of $32$ elements.
The corresponding tensor products of Pauli operators can be grouped into four sets: 
\begin{eqnarray}
\{XXY \openone,XY\openone X,X\openone XY, \openone YYY,\dots\}, \nonumber \\ 
\{XYX\openone,YY\openone Y,Y\openone XX,\openone XXY,\dots\}, \nonumber \\ 
\{YXX\openone,XX\openone Y,Y\openone YY,\openone XYX,\dots\}, \nonumber \\ 
\{YYY\openone,YX\openone X,X\openone YX,\openone YXX,\dots\}, \nonumber 
\end{eqnarray}
where the dots denote four more operators being 
the previous four operators with $X$ replaced by $Y$ and vice versa.
All operators in each set anti-commute, therefore the bound is proved.

To give a concrete example of monogamy of a well-known inequality
we choose the inequality due to Mermin \cite{MERMIN}:
$E_{112} + E_{121} + E_{211} - E_{222} \le 2$,
where $E_{klm}$ denote the correlation functions.
Since the classical bound of the Mermin inequality is $2$,
and not $1$ as we have assumed in our derivation,
the new "Mermin monogamy" is $\mathcal{M}_{ABC}^2+\mathcal{M}_{ABD}^2+\mathcal{M}_{ACD}^2+\mathcal{M}_{BCD}^2 \leq 16$, where $\mathcal{M}$ is the quantum value of the corresponding Mermin parameter.
The bound of the new monogamy relation can be achieved in many ways.
If a triple of observers share the GHZ state, they can obtain maximal violation of $4$
and the remaining triples observe vanishing Mermin quantities $\mathcal{M}$.
This can be attributed to maximal entanglement of the GHZ state.
It is also possible for two and three triples to violate Mermin inequality non-maximally, and at the same time to achieve the bound. 
For example, the state $\frac{1}{2}\left(|0001\rangle+|0010\rangle +i\sqrt{2}|1111\rangle\right)$ allows $ABC$ and $ABD$ to obtain $\mathcal{M} = 2\sqrt{2}$, 
and the state $\frac{1}{\sqrt{6}}\left(|0001\rangle+|0010\rangle+|0100\rangle+i\sqrt{3}|1111\rangle\right)$ allows $ABC$, $ABD$ and $ACD$ to obtain $\mathcal{M} = \tfrac{4}{\sqrt{3}}$.
Note that it is impossible to violate all four inequalities simultaneously.

We now derive new monogamy relations for $N$ qubits.
Consider scenario of Fig. \ref{FIG_GRAPHS}c,
in which $N$ is odd, $A$ is the fixed qubit and the remaining $N-1$ qubits are split into two groups $\vec B = (B_1,...,B_M)$ and $\vec C = (C_1,...,C_M)$ each containing $M=\tfrac{1}{2}(N-1)$ qubits.
We shall derive the trade-off relation between violation of $(M+1)$-partite Bell inequality by parties $A \vec B$ and $A \vec C$.
Using condition (\ref{UP_BOUND}),
the elements of the correlation tensor which enter the bound of $\mathcal{L}_{A \vec B}^2  + \mathcal{L}_{A \vec C}^2$
are of the form $T_{k l_1 \dots l_M 0 \dots 0}$ and $T_{k 0 \dots 0 m_1 \dots m_M}$.
The corresponding Pauli operators can be arranged into $2^{M}$ sets of four mutually anti-commuting operators each:
$\vec A_{1S}  = \{XXS I, XYSI, YIXS, YIYS\}$,
$\vec A_{2S}  = \{YXS I, YYS I,  XIXS,  XIYS\}$,
where $S$ stands for all $2^{M-1}$ combinations of $X$'s and $Y$'s for $M-1$ parties,
and $I = \openone^{\otimes M}$ is identity operator on $M$ neighboring qubits.
Therefore, according to the theorem, we arrive at the following trade-off: $\mathcal{L}_{A \vec B}^2 + \mathcal{L}_{A \vec C}^2  \le 2^M$.

The bound of this inequality is tight in the sense that there exist quantum states
achieving the bound for all allowed values of $\mathcal{L}_{A \vec B}$ and $\mathcal{L}_{A \vec C}$.
This is a generalization of a similar property for CHSH monogamy \cite{TV2006}.
The state of interest can be chosen as
\begin{equation}
|\psi \rangle = \tfrac{1}{\sqrt{2}} \cos \alpha \left( | 0 \vec 0 \vec 0 \rangle + |1 \vec 0 \vec 1 \rangle \right) + \tfrac{1}{\sqrt{2}} \sin \alpha \left( | 1 \vec 1 \vec 0 \rangle + |0 \vec 1 \vec 1 \rangle \right),
\label{PSI_MONO}
\end{equation}
where e.g. $|1 \vec 0 \vec 1 \rangle$ denotes a state in which qubit $A$ is in the $| 1 \rangle$ eigenstate of local $Z$ basis,
all qubits of $\vec B$ are in state $| 0 \rangle$ of their local $Z$ bases, and all qubits of $\vec C$ are in state $| 1 \rangle$ of their respective $Z$ bases.
The non-vanishing correlation tensor components in $xy$ plane,
which involve only $(M+1)$-partite correlations are
$T_{x \vec w \vec 0} = \pm \sin 2\alpha$,
$T_{x \vec 0 \vec w} = \pm 1$, and 
$T_{y \vec 0 \vec v} = - \cos 2\alpha$,
where $\vec w$ contains even number of $y$ indices, other indices being $x$,
and $\vec v$ contains odd number of $y$ indices, other indices again being $x$.
There are $\sum_{k=1}^{\lfloor M/2 \rfloor} {M \choose 2 k} = 2^{M-1}$
correlation tensor elements of each type and consequently
\begin{equation}
\mathcal{L}_{A \vec B}^2 = 2^{M-1} \sin^2 2\alpha, \quad \mathcal{L}_{A \vec C}^2 = 2^{M-1}(1+ \cos^2 2\alpha).
\label{TIGHT}
\end{equation}
Therefore, the bound is always achieved
and all allowed values of $\mathcal{L}_{A \vec B}$ and $\mathcal{L}_{A \vec C}$
can be attained either by the state (\ref{PSI_MONO}) or the state with the role of qubits $\vec B \leftrightarrow \vec C$ interchanged. The underlying reason why the above trade-off allows for violation by both $A \vec B$ and $A \vec C$
is the fact that sets of anti-commuting operators of the Bell parameters can contain at most four elements.

Now we present a much stronger new monogamy related to the graph in Fig. \ref{FIG_GRAPHS}d.
Consider $M$-partite Bell inequalities corresponding to different paths from the root of the graph to its leaves ($M=3$ in Fig. \ref{FIG_GRAPHS}d).
There are $2^{M-1}$ such inequalities and we shall prove that their quantum mechanical values obey
\begin{equation}
\mathcal{L}_1^2 + \dots + \mathcal{L}_{2^{M-1}}^2\le 2^{M-1},
\label{STRONG}
\end{equation}
where $\mathcal{L}_j$ is the quantum value for the $j$-th Bell parameter in the graph.
To prove this, we construct $2^{M-1}$ sets of anti-commuting operators,
each set containing $2^M$ elements, such that they exhaust all 
correlation tensor elements which enter the bound of the left-hand side of (\ref{STRONG})
after application of condition (\ref{UP_BOUND}).
The construction also uses the graph of the binary tree.
We begin at the root, to which we associate a set of two anti-commuting operators, $X$ and $Y$, for the corresponding qubit.
A general rule now is that if we move up in the graph from qubit $A$ to qubit $B$ 
we generate two new anti-commuting operators by placing $X$ or $Y$ at position $B$ to the operator which had $X$ at position $A$.
Similarly, if we move down in the graph to qubit $C$ we generate two new anti-commuting operators by placing $X$ or $Y$ at position $C$ to the operator which contained $Y$ at position $A$.
For example, starting from the set of operators $(X,Y)$ by moving up we obtain $(XX\openone,XY\openone)$, and by moving down we have $(Y \openone X ,Y \openone Y)$. The next sets of operators are $(XX\openone X\openone \openone \openone, XX\openone Y\openone \openone \openone)$, $( XY\openone\openone X \openone \openone, XY\openone \openone Y \openone \openone)$, $(Y\openone X\openone \openone X\openone, Y\openone X \openone \openone Y \openone)$ and $(Y \openone Y \openone \openone \openone X, Y \openone Y\openone \openone \openone Y)$ if we move from the root: up up, up down, down up and down down, respectively.
By following this procedure in the whole graph we obtain a set of $2^M$ mutually anti-commuting operators.
According to this algorithm the anti-commuting operators can be grouped in pairs having the same Pauli operators except for the qubits of the last step (the leaves of the graph).
There are $2^{M-1}$ such pairs corresponding to distinct combinations of tensor products of $X$ and $Y$ operators on $M-1$ positions.
Importantly, in different operators these positions are different and to generate the whole set of operators entering 
the bound we have to perform suitable permutations of positions. 
Such permutations always exist and they do not affect anti-commutativity.
Finally we end up with the promised $2^{M-1}$ sets of $2^{M}$ anti-commuting operators each,
which according to Eq. (\ref{INEQ}) give the bound of (\ref{STRONG}).

The inequality (\ref{STRONG}) is stronger than the previous trade-off relation in the sense that it does not allow 
simultaneous violation of all the inequalities of its left-hand side.
All other patterns of violations are possible as we now show.
Choose any number, $m$, of Bell inequalities, i.e. paths in the Fig. \ref{FIG_GRAPHS}d.
Altogether they involve $n$ parties which share the following quantum state
\begin{equation}
\ket{\psi_n} = \frac{1}{\sqrt{2}} | \underbrace{0 \dots 0}_{n} \rangle + \frac{1}{\sqrt{2 m}} \sum_{j = 1}^m | 0 \dots 0 \underbrace{1 \dots 1}_{\mathcal{P}_j} 0 \dots 0 \rangle,
\end{equation}
where $\mathcal{P}_j$ denotes parties involved in the $j$-th Bell inequality.
Note that all states under the sum are orthogonal as they involve different parties.
The only non-vanishing components of the correlation tensor of this state
have even number of $y$ indices for the parties involved in the Bell inequalities.
Squares of all these components are equal to $\tfrac{1}{m}$ which gives $\mathcal{L}_j^2 = \tfrac{2^{M-1}}{m}$ for each Bell inequality $j=1,\dots, m$.
Therefore, all $m$ Bell inequalities are violated as soon as $m < 2^{M-1}$.
Moreover, the sum of these $m$ Bell parameters saturates the bound of (\ref{STRONG})
and therefore independently of the state shared by other parties
the remaining Bell parameters of (\ref{STRONG}) all vanish.

In conclusion, we have derived monogamy of multipartite Bell inequality violations
which are all quadratic functions of Bell parameters.
As such these relations are stronger than those following from no-signaling principle alone,
which are linear in Bell parameters \cite{BLMPSR2005,MAN2006,TONER2009,PB2009}.
Indeed, most of our monogamies are tight in the sense that they precisely identify the set of Bell violations allowed by quantum theory.
Our proofs are within quantum formalism and utilize the bounds imposed by the complementarity principle. These bounds were established for dichotomic observables and are applicable to any Bell inequality involving these, it would be useful to extend the formalism to measurements with more outcomes. It would also be interesting to see if the Bell violation trade-offs can be derived without using quantum formalism, a candidate for this task is the principle of information causality \cite{IC}.

\emph{Acknowledgements}.
This research is supported by the National Research Foundation and Ministry of Education in Singapore.
WL is supported by the EU program Q-ESSENCE (Contract No.248095),  the  MNiSW Grant no. N202 208538 and by the Foundation for Polish Science.

\end{document}